\documentclass[aps,prb,citeautoscript,superscriptaddress,preprint,onecolumn]{revtex4}

\usepackage{graphicx}
\usepackage{amsfonts,amsmath,amssymb,amsthm}
\usepackage{epstopdf}
\usepackage{upgreek,xspace}
\usepackage{chngcntr}
\usepackage{xr}
\usepackage[colorlinks,allcolors=blue]{hyperref}
\usepackage[version=3]{mhchem} 

\begin{document}
\bibliographystyle{apsrev}
\title{Stacking-Selective Epitaxy of Rare-Earth Diantimonides}
\author{Reiley~Dorrian}
\affiliation{Department of Applied Physics and Materials Science, California Institute of Technology, Pasadena, California 91125, USA.}

\author{Jinwoong~Kim}
\affiliation{Department of Physics and W. M. Keck Computational Materials Theory Center, California State University, Northridge, Northridge,
California 91330, USA.}

\author{Adrian~Llanos}
\affiliation{Department of Applied Physics and Materials Science, California Institute of Technology, Pasadena, California 91125, USA.}

\author{Veronica~Show}
\affiliation{Department of Applied Physics and Materials Science, California Institute of Technology, Pasadena, California 91125, USA.}

\author{Mizuki~Ohno}
\affiliation{Department of Applied Physics and Materials Science, California Institute of Technology, Pasadena, California 91125, USA.}

\author{Nicholas~Kioussis}
\affiliation{Department of Physics and W. M. Keck Computational Materials Theory Center, California State University, Northridge, Northridge,
California 91330, USA.}

\author{Joseph~Falson}
\email{falson@caltech.edu}
\affiliation{Department of Applied Physics and Materials Science, California Institute of Technology, Pasadena, California 91125, USA.}
\affiliation{Institute for Quantum Information and Matter, California Institute of Technology, Pasadena, California 91125, USA.}

\begin{abstract}
Deterministic control of the layering configuration of two-dimensional quantum materials plays a central role in studying their emergent electronic properties. Here we demonstrate \textit{in-situ} control over competing stacking configurations in thin film crystals of the rare-earth diantimonides by synthesizing in proximity to competing structural orders. A crossover between the epitaxially stabilized monoclinic structure and the orthorhombic structure commonly observed in bulk crystals is navigated through three axes; the relative cation/anion ratio, growth temperature, and choice of lanthanide ion, culminating with a comparative magnetotransport study of single-yet-distinct phase \ce{CeSb2} films. These results set the stage for an expanded search for hidden stacking configurations in layered compounds which have evaded detection. 
\end{abstract}

\maketitle

An expansive parameter space for studying emergent electronic phases in low dimensional materials can be realized when considering the stacking degree of freedom between sheets of layered compounds\cite{zhang:2011,cao:2018,park:2023,lu:2024}. In some classes of materials, such as layered intermetallics, there naturally exists an array of stacking configurations which are stabilized based on the precise chemical details of the crystal system.\cite{ren:2009,seibel:2015,sinha:2021,hart:2023,hu:2024} One example is the \ce{\textit{Ln}Sb2} ($Ln$ = lanthanide element) family. Comprised of sequential \textit{Ln}-Sb corrugated layers separating square-nets of Sb,\cite{wang:1967} distinct stacking arrangements are achieved by shearing layers in the $a$-$b$ plane and through dimerization of intralayer atoms. \cite{llanos:2024} These materials display numerous distinct structure-types, discussed below, due to the variation in electron count as the \textit{Ln} site is tuned across the lanthanide period.\cite{hulliger:1978,tremel:1987,papoian:2000,ohno:2021} Several of these structures show instabilities under pressure evidenced by hysteretic temperature dependent transport features accompanied by qualitative changes in the electronic ground states, such as the emergence of superconductivity concomitant to the suppression of magnetic order.\cite{kagayama:2000,weinberger:2023,squire:2023}

The recent discovery of \ce{LaSb2} taking a previously uncharacterized structure-type when grown as thin films has highlighted the potential for discovering new stacking configurations which have evaded detection.\cite{llanos:2024} The crystals grown using molecular beam epitaxy (MBE) were characterized by a monoclinic shear and compressed $c$-axis relative to the \ce{SmSb2} structure observed in flux grown bulk crystals, and displayed a superconducting critical temperature exceeding bulk crystals by a factor of two. Additionally, no hysteretic transport features indicative of a phase instability were found in the temperature dependence of the resistivity. Perhaps most enigmatic, first principles calculations revealed that the previously uncharacterized stacking sequence was indeed the lowest energy configuration, putting the widely observed \ce{SmSb2} structure as an excited configuration. Here, we identify electron counting within the crystal structure as a prominent effect in dictating the structure-type which forms as thin films. We use \ce{CeSb2} as our platform to explore this point; growing at high temperatures which deprives the crystal of antimony results in the novel monoclinic stacking configuration, while antimony rich conditions produce the bulk \ce{SmSb2} structure-type. This approach culminates in the ability to execute polymorph-selective epitaxy of the \ce{CeSb2} system while probing the chemical environments which control the competing stacking configurations. We directly compare the electronic transport properties of both structures, demonstrating the interplay between the structural and electronic degrees of freedom in this class of layered quantum materials.

\begin{figure*}[ht]
    \centering
    \includegraphics[width=80mm]{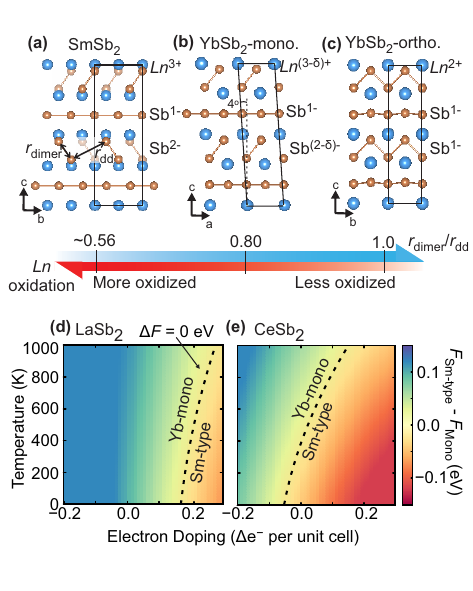}
    \caption{Schematic view of the (a) Sm-type, (b) novel Yb-mono, and (c) Yb-type orthorhombic polymorphs. Parametrizing each structure by the relative intra- to inter-dimer spacing of Sb anions in the ionic spacer layers provides a proxy for the valence state of the rare-earth cation. Temperature versus electron-doping phase diagrams for (d) \ce{LaSb2} and (e) \ce{CeSb2} as predicted by DFT.}
    \label{Fig1}
\end{figure*}

We begin with a discussion on electron counting within the unit cell. The competing layering configurations relevant to this study are laid out schematically in Fig.~\ref{Fig1}. The well characterized stacking configurations identified in bulk crystals are stabilized by predominantly $3+$ (Sm-type) and $2+$ (Yb orthorhombic-type, noted as Yb-ortho)\cite{tremel:1987} oxidation states on the rare-earth ion. Close inspection of the bonding geometry of the hypervalent Sb network allows the establishment of a quantitative proxy of the $Ln$-site valence state. The logic follows the framework developed by Papoian and Hoffmann;\cite{papoian:2000} a \ce{Sb^{1-}} valence state is typically associated with square planar or zig-zag bonding geometries, while \ce{Sb^{2-}} results in dimerization. All three structure-types in Fig.~\ref{Fig1} share the Sb square-net motif yet the geometry in their \ce{\textit{Ln}Sb} spacer layers differ; dimerization between neighboring spacer-layer Sb atoms is evident in the Sm-type structure (Fig.~\ref{Fig1}a) while a zig-zag chain forms in the Yb-type stacking (Fig.~\ref{Fig1}c). The ratio of the intra- to inter- dimer distance ($r_\text{dimer}/r_\text{dd}$) quantifies the degree of Sb-Sb dimerization within the network and varies from $\approx 0.56$ in the Sm-type structure to $1$ in the 1D chains of the Yb-type. In contrast, the novel Yb monoclinic structure (noted as Yb-mono) \cite{llanos:2024} falls in the middle of these two known structure-types, at $r_\text{dimer}/r_\text{dd}\approx$ 0.80. This suggests that an intermediate valence state below full trivalence may be realized on the $Ln$-site in the Yb-mono structure, as shown in Fig.~\ref{Fig1}b. Note that while low-valence lanthanides are generally unstable in the light rare-earths, their discovery in organometallic molecular complexes\cite{hitchcock:2008} suggests that a fractional (3$-\delta$)+ oxidation state may be possible in an extended metal network such as our system. We hypothesize that the oxidation pressure applied by the Sb anion and growth temperature are partially responsible for this effect, which we will explore below. 

To guide the experimental efforts, we have performed first-principles calculations of the free energy difference ($\Delta F $) between Sm-type and Yb-mono structures as a function of electron doping and temperature in Figs.~\ref{Fig1}d,e. Elevated temperatures are considered due to phononic entropy contributions to the free energy. In both systems electron doping favors stabilization of the Sm-type phase, while high temperatures favor the Yb-mono phase. We find that in both \ce{LaSb2} and \ce{CeSb2} there exists a crossover between the Sm-type and Yb-mono structures, but this only occurs at zero electron doping in \ce{CeSb2} at finite temperatures.

\begin{figure*}[ht]
    \centering
    \includegraphics[width=160mm]{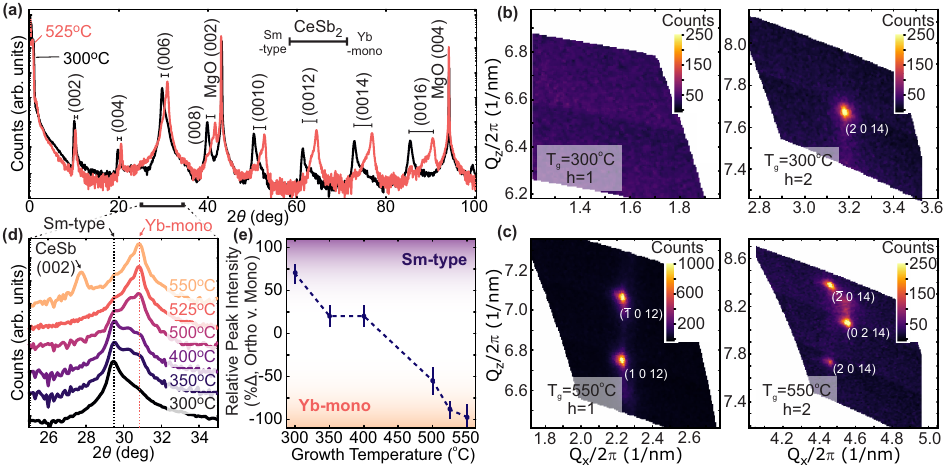}
    \caption{(a) XRD diffraction patterns from \ce{CeSb2} films grown at distinct substrate temperatures. Reciprocal space maps taken around the \{1 0 12\} and \{2 0 14\} diffraction conditions identify the two films' structure types as (b) $T_\mathrm{g}=300^\circ$C, Sm-type and (c) $T_\mathrm{g}=525^\circ$C, Yb-mono. (d) Expanded view of (0 0 6) Bragg peak for a series of growth temperatures. These Bragg peaks are fit to a two-gaussian lineshape, quantifying the relative diffraction intensity of each polyorph as a function of growth temperature (e).}
    \label{Fig2}
\end{figure*}

Figure \ref{Fig2} summarizes the X-ray diffraction (XRD) results of a series of films grown at increasing growth temperature ($T_\mathrm{g}$) under otherwise identical growth conditions. The out-of-plane diffraction patterns for $T_\mathrm{g} = 300^\circ$C and $525^\circ$C in Fig.~\ref{Fig2}a show reflections from the (001) \ce{CeSb2} lattice planes without any impurity phases evident. The two distinct sets of periodic peaks correspond to out-of-plane lattice constants of $18.16$ \AA~($T_\mathrm{g} = 300^\circ$C) and $17.38$ \AA~($T_\mathrm{g} = 525^\circ$C). Such large $d$-spacings are similar to the $c$-axis lattice parameters of bulk Sm-type \ce{CeSb2} for low $T_\mathrm{g}$ and \ce{Yb}-mono \ce{CeSb2} (approximated by a uniform lanthanide contraction of $\approx 2.3\%$ applied to Yb-mono \ce{LaSb2}) for high $T_\mathrm{g}$ \cite{wang:1967,llanos:2024}. We definitively identify these structures by asymmetric reciprocal space mapping (RSM) along the $h=1$ (Fig.~\ref{Fig2}b) and $h=2$ (Fig.~\ref{Fig2}c) rods for both $T_\mathrm{g}$. The $T_\mathrm{g}=525^\circ$C film shows twinned spots split along $Q_z$ for both the $(\pm1 \; 0\; 12)$ and $(\pm2\; 0\; 14)$ film peaks, indicating a monoclinic tilt in the direction of the $a$ lattice vector.\cite{llanos:2024} Meanwhile, the $T_\mathrm{g}=300^\circ$C film shows a single diffraction spot for $h=2$ and no diffraction for the $(1\;0\;12)$ series, consistent with no monoclinic tilt and the extinction conditions for the $Cmca$ space group (SG no. 64). Analysis of these RSM film peaks returns in-plane lattice constants $a = 6.27$ \AA\, and $b = 6.16$ \AA\, for the $300^\circ$C film and $a=4.49$ \AA, $b=4.39$ \AA, and $\beta = 86.10^\circ$ for the $525^\circ$C film (tabulated in Supplementary Materials). These observations confirm a crossover from the bulk Sm-type structure to the Yb-mono structure as $T_\mathrm{g}$ is increased.

The transition between these structural phases is observed in Fig.~\ref{Fig2}d by zooming in on the (006) diffraction condition. Both phases clearly coexist for $T_\mathrm{g}$ between $300^\circ$C and $550^\circ$C as seen by the pronounced signal at both $2\theta \approx 29.5^o$ and $30.8^o$. A CeSb peak appears at $T_\mathrm{g}\approx 550^\circ$C and above due to thermal desorption of volatile Sb from the substrate surface during growth. Figure~\ref{Fig2}e quantifies the crossover between the phases by fitting the measured (006) Bragg peaks (on a logarithmic scale) to a superposition of two Gaussian lineshapes, one centered on each phases' diffraction angle (see Fig. S5). The percent-difference between the resulting amplitude of the two Gaussians provides a quantitative approximation of relative phase concentration, in which we see a monotonic trend towards the Yb-mono phase as a function of rising $T_\mathrm{g}$. Elevating the temperature in experiment results in both Sb desorption as well phononic entropy contributions, and while only the latter is captured in the calculations in Fig.~\ref{Fig1}d,e, these results appear qualitatively consistent with the thermodynamics of undoped \ce{CeSb2}.

It is worth noting that the logarithmic scale in Figure~\ref{Fig2}d amplifies the asymmetry of the Bragg peaks, which in turn affects the relative peak intensities of Figure~\ref{Fig2}e. This should therefore be taken only as a proxy for the phase purity of each film, which is more clearly observed on a linear-scale in Figure S5. Furthermore, the third peak-like feature seen around $2\theta \approx 30^o$ has not been associated with any known or predicted structure within the Ce-Sb system. Instead, we hypothesize that it is the effect of stacking disorder within the mixed-configuration films. Stacking faults in layered samples with multiple competing polymorphs are known to result in either additional XRD reflections \cite{pujar:1995} or peak shifting to lower $2\theta$.\cite{kopp:2012,wagner:2014} A slight shift of the Yb-mono peak due to layering disorder in the presence of the competing Sm-type would explain the observed feature.

\begin{figure*}[ht]
    \centering
    \includegraphics[width=80mm]{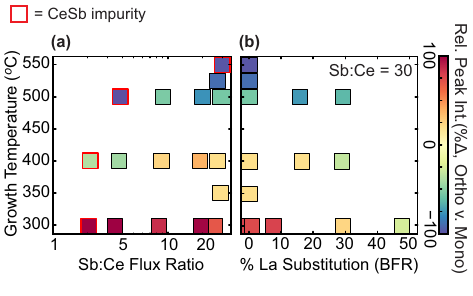}
    \caption{Polymorph-stability phase diagram expanded as a function of (a) experimental Sb:Ce beam flux ratio and (b) atomic-percent substitution of La ions in place of Ce, as estimated by the provided La:Ce beam flux ratio. Phase purity between the Sm-type and Yb-mono polymorphs is quantified by two-gaussian fits to the (0 0 6) peak as in Figure~2e and represented by hue. Films with measureable CeSb impurity phases are flagged with a red outline.}
    \label{Fig3}
\end{figure*}

Figure~\ref{Fig3}a explores the effects of varying the Sb:Ce beam flux ratio from $2 \sim 30$ across a range of growth temperatures. The color coded data points quantify the ratio of peak intensity of the competing phases, as discussed in Fig.~\ref{Fig2}e. At $T_\mathrm{g} = 300^\circ$C, the Sm-type phase is generally favored even when reducing Sb flux significantly. At $T_\mathrm{g} = 400^\circ$C, however, these phases coexist and it is possible to resolve a tendency towards the Yb-mono phase with decreasing Sb flux. At even higher temperatures, the Yb-mono phase begins to dominate for all flux ratios. A qualitative resemblance is immediately apparent between the data in Fig.~\ref{Fig1} and Fig.~\ref{Fig3}a; the amount of Sb overpressure here plays the role of electron-doping, where an induced Sb deficiency is analogous to hole-doping as the total number of valence electrons in the unit cell is decreased.  This effect is converse from the perspective of the Ce cation oxidation state. In growth conditions which are deficient in the electronegative Sb, the rare-earth ion experiences a decreased oxidation pressure and is therefore more easily stabilized in the slightly reduced (3$-\delta$)+ valence required to form the Yb-mono structure. In highly Sb-rich conditions it is increasingly probable that the more oxidized 3+ state is formed, thereby stabilizing the Sm-type polymorph.

We further explore this through La substitution on the Ce site in Fig.~\ref{Fig3}b under a fixed Sb:Ce flux ratio of 30. According to Figs.~\ref{Fig1}d,e, La substitution can similarly drive a transition from the Sm-type to the Yb-mono phase for a wide range of temperatures and electron doping ranges. The data in Fig.~\ref{Fig3}b show that for all temperatures the substitution of \ce{Ce} with \ce{La} drives the system towards the Yb-mono phase.

We now discuss how the experimental setting for crystal growth can affect the phase stability observed. Given that both \ce{LaSb2} and \ce{CeSb2} are consistently observed in the Sm-type structure in bulk reports, the discussions above imply that bulk crystals are effectively electron-doped with the $Ln$-site having a valence close to 3+. This seems plausible given the use of self-flux for single crystal growth which provides ample supply of the Sb anion. Epitaxial crystals are meanwhile grown under high vacuum conditions with continual desorption of volatile species from the substrate surface. Thus, it is possible to access both antimony rich and poor growth conditions in epitaxy, which may serve as a knob for tuning the oxidation pressure on the $Ln$-site. The wide range of temperatures available in epitaxy also permits insight into the role of phonon entropy. And while \ce{LaSb2} agrees well with the predicted phase stability in undoped conditions, epitaxial \ce{CeSb2} allows us to study the crossover between the two polymorphs as the growth conditions are tuned experimentally.

\begin{figure*}[ht]
    \centering
    \includegraphics[width=170mm]{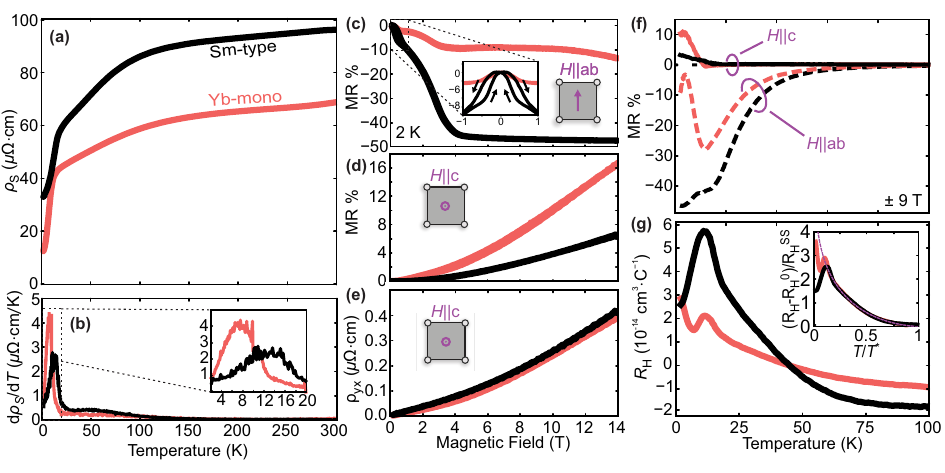}
    \caption{(a) Temperature-dependent longitudinal resistivity of both polymorphs, alongside (b) the derivative with respect to temperature. magnetoresistance ratio for (c) in-plane and (d) out-of-plane magnetic field and (d) Hall resistivity at 2 K. (f) Magnitude of magnetoresistance ratio at $|H| = 9$ T for both field configurations and (g) approximate Hall coefficient as a function of temperature. Inset: temperature-dependent Hall coefficient fit to two-fluid skew scattering model.}
    \label{Fig4}
\end{figure*}

The ability to selectively stabilize both polymorphs of \ce{CeSb2} as epitaxial films allows us to directly compare their electronic transport properties. Figure~\ref{Fig4} summarizes this comparison via temperature-dependent sheet resistivity $\rho_\text{S}(T)$ (Figs.~\ref{Fig4}a,b), magnetoresistance (Figs.~\ref{Fig4}c,d,f), and Hall effect (Figs.~\ref{Fig4}e,g) measurements performed on 30 nm films of each structure. Both phases show metallic transport with $\rho_\text{S}(T)$ resembling that of bulk crystals and other Kondo-lattice compounds.\cite{canfield:1991,budko:1998, kagayama:2000, coleman:2015} The films' residual resistivity ratios (RRR) are found to be $\approx 2.9$ for the Sm-type film and $\approx 5.5$ for the Yb-mono film. Both films see a downturn in resistivity centered around 100-150$^\circ$C as is often observed in Ce-based heavy fermion compounds.\cite{nishida:2006,zhang:2024,posey:2024} The sharper drop in resistivity below $T$=20~K marks the onset of magnetic ordering and is found to be comparatively larger in amplitude and lower in temperature for the Yb-mono phase, indicating a slight suppression of magnetic order in the lower-symmetry phase alongside a reduction in spin-disorder scattering.

The field-dependent magnetoresistance MR(\%)=$100\times\frac{\rho_\text{S}(H)-\rho_\text{S}(H=0\text{ T})}{\rho_\text{S}(H=0\text{ T})}$ for both in-plane and out-of-plane magnetic fields are presented in Figs.~\ref{Fig4}c-d alongside the Hall resistivity in Fig.~\ref{Fig4}e, all collected at 2 K. Clear meta-magnetic transitions are seen for both films as evidenced by drops in resistivity under an increasing in-plane field, indicating several low-$T$ magnetic phases as reported in bulk.\cite{zhang:2017,trainer:2021} A magnetic ground state with in-plane easy axis is evidenced by hysteretic behavior for $|H|<1$ T, followed by a pair of relatively broad transitions centered around $\approx 1$ T and $3$ T. These observations are consistent with magnetoresistance reports in bulk crystals, where the transitions are suggested to represent a sudden spin-flip and a more gradual spin-rotation, respectively, of the localized Ce moments as the system traverses a complicated magnetic phase diagram.\cite{budko:1998, kagayama:2000} For the Sm-type film, the resistivity falls to nearly 50\% of its zero-field value after this second spin-rotation and largely saturates up to 14 T, indicative of a fully polarized paramagnetic state. The Yb-mono film, on the other hand, shows only about 10\% negative MR after the second metamagnetic transition and begins another gradual downturn at fields above 10 T. This suggests an incomplete saturation of the spins which requires much higher fields to fully polarize: it is possible that the Yb-mono structure results in more geometric frustration for the Ce sites than the Sm-type structure, allowing spin-disorder scattering to persist up to much higher in-plane fields. In contrast, no clear signs of magnetic ordering are observed for the out-of-plane field. We are careful not to make quantitative conclusions about the carrier densities through magnetotransport in materials with magnetic ordering and Kondo effects \cite{nair:2012}. Nonetheless, the positive sign and curvature of the Hall resistivity observed in both films are reminiscent of that seen in \ce{LaSb2} films, suggesting at least two carrier types with conduction dominated by the hole species. The positive, roughly quadratic magnetoresistance likely arises from the orbital scattering of these carriers as described by the Drude model. 

Figures~\ref{Fig4}f-g summarize the temperature-dependence of magnetoresistance ratio at $|H|=9$ T and Hall coefficient $R_\text{H}$=$\frac{\rho(H=9\text{ T})-\rho(H=- 9\text{ T})}{18 \text{ T}}$ up to 100 K. The increase of $R_\text{H}$ with decreasing temperature is a well-documented characteristic of Kondo lattice systems, where skew-scattering of the conduction electrons by localized Ce moments becomes significant as the system approaches the Kondo coherence temperature $T_\text{coh}$.\cite{holter:1985,fert:1987,hamzic:1988, nair:2012} A phenomenological model has been put forth to interpret this behavior in terms of a two-fluid picture, wherein a heavy Kondo liquid begins to form from the standard Fermi liquid at a characteristic temperature $T^*$.\cite{zhang:2017} This gives 
\begin{gather}
    R_\text{H} = R_H^\mathrm{SS}(1-T/T^*)^{3/2}(1+\log(T^*/T)) +R_\mathrm{H}^0
\end{gather}
(where $R_H^\mathrm{SS}$ weights the skew-scattering contribution and $R_\mathrm{H}^0$ is the ordinary Hall coefficient, taken to be constant) which we fit to our data in the inset. Below the Kondo coherence temperature the skew-scattering becomes coherent and its contribution to $R_\text{H}$ precipitously drops, resulting in a local maximum in $R_\text{H}$ centered around $T_\text{coh}$. This behavior is observed in both films with a common coherence temperature of about 12~K. The more dramatic peak structure seen in the Sm-type film may indicate stronger hybridization between conduction and localized electrons in the structure.

In summary, we have demonstrated experimental control over competing stacking configurations of \ce{\textit{Ln}Sb2} crystals grown as thin films. Bulk Sm-type and novel Yb-mono polymorphs are selectable in \ce{CeSb2} by tuning the flux ratio and substrate temperature during growth, and a crossover between these structures is supported by first-principles calculations. The unique ability to work across a wide range of Sb dosing conditions is revealed as important in driving the phase transition, which is inaccessible in conventional flux bulk synthesis. These results highlight the importance of thin film growth in discovering new structures in quantum materials otherwise inaccessible by traditional bulk-synthesis methods.

\section*{Acknowledgments}
We appreciate discussions with Bharat Jalan. This material is based upon work supported by the National Science Foundation Graduate Research Fellowship under Grant No. 2139433. We acknowledge funding provided by the Air Force Office of Scientific Research (Grant number FA9550-22-1-0463), the Gordon and Betty Moore Foundation’s EPiQS Initiative (Grant number GBMF10638), and the Institute for Quantum Information and Matter, a NSF Physics Frontiers Center (NSF Grant PHY-2317110). We also acknowledge the Beckman Institute for their support of the X-ray Crystallography Facility at Caltech.

\section*{Conflict of Interest}
The authors declare no conflict of interests.

\newpage

\newpage

\newpage

\newpage

\clearpage 

\section*{}
\bibliography{bib}

\begin{thebibliography}{42}
\expandafter\ifx\csname natexlab\endcsname\relax\def\natexlab#1{#1}\fi
\expandafter\ifx\csname bibnamefont\endcsname\relax
  \def\bibnamefont#1{#1}\fi
\expandafter\ifx\csname bibfnamefont\endcsname\relax
  \def\bibfnamefont#1{#1}\fi
\expandafter\ifx\csname citenamefont\endcsname\relax
  \def\citenamefont#1{#1}\fi
\expandafter\ifx\csname url\endcsname\relax
  \def\url#1{\texttt{#1}}\fi
\expandafter\ifx\csname urlprefix\endcsname\relax\def\urlprefix{URL }\fi
\providecommand{\bibinfo}[2]{#2}
\providecommand{\eprint}[2][]{\url{#2}}

\bibitem[{\citenamefont{Zhang et~al.}(2011)\citenamefont{Zhang, Jung, Fiete, Niu, and MacDonald}}]{zhang:2011}
\bibinfo{author}{\bibfnamefont{F.}~\bibnamefont{Zhang}}, \bibinfo{author}{\bibfnamefont{J.}~\bibnamefont{Jung}}, \bibinfo{author}{\bibfnamefont{G.~A.} \bibnamefont{Fiete}}, \bibinfo{author}{\bibfnamefont{Q.}~\bibnamefont{Niu}}, \bibnamefont{and} \bibinfo{author}{\bibfnamefont{A.~H.} \bibnamefont{MacDonald}}, \bibinfo{journal}{Physical Review Letters} \textbf{\bibinfo{volume}{106}}, \bibinfo{pages}{156801} (\bibinfo{year}{2011}), \bibinfo{note}{publisher: American Physical Society}, \urlprefix\url{https://link.aps.org/doi/10.1103/PhysRevLett.106.156801}.

\bibitem[{\citenamefont{Cao et~al.}(2018)\citenamefont{Cao, Fatemi, Fang, Watanabe, Taniguchi, Kaxiras, and Jarillo-Herrero}}]{cao:2018}
\bibinfo{author}{\bibfnamefont{Y.}~\bibnamefont{Cao}}, \bibinfo{author}{\bibfnamefont{V.}~\bibnamefont{Fatemi}}, \bibinfo{author}{\bibfnamefont{S.}~\bibnamefont{Fang}}, \bibinfo{author}{\bibfnamefont{K.}~\bibnamefont{Watanabe}}, \bibinfo{author}{\bibfnamefont{T.}~\bibnamefont{Taniguchi}}, \bibinfo{author}{\bibfnamefont{E.}~\bibnamefont{Kaxiras}}, \bibnamefont{and} \bibinfo{author}{\bibfnamefont{P.}~\bibnamefont{Jarillo-Herrero}}, \bibinfo{journal}{Nature} \textbf{\bibinfo{volume}{556}}, \bibinfo{pages}{43} (\bibinfo{year}{2018}), ISSN \bibinfo{issn}{1476-4687}, \bibinfo{note}{publisher: Nature Publishing Group}, \urlprefix\url{https://www.nature.com/articles/nature26160}.

\bibitem[{\citenamefont{Park et~al.}(2023)\citenamefont{Park, Cai, Anderson, Zhang, Zhu, Liu, Wang, Holtzmann, Hu, Liu et~al.}}]{park:2023}
\bibinfo{author}{\bibfnamefont{H.}~\bibnamefont{Park}}, \bibinfo{author}{\bibfnamefont{J.}~\bibnamefont{Cai}}, \bibinfo{author}{\bibfnamefont{E.}~\bibnamefont{Anderson}}, \bibinfo{author}{\bibfnamefont{Y.}~\bibnamefont{Zhang}}, \bibinfo{author}{\bibfnamefont{J.}~\bibnamefont{Zhu}}, \bibinfo{author}{\bibfnamefont{X.}~\bibnamefont{Liu}}, \bibinfo{author}{\bibfnamefont{C.}~\bibnamefont{Wang}}, \bibinfo{author}{\bibfnamefont{W.}~\bibnamefont{Holtzmann}}, \bibinfo{author}{\bibfnamefont{C.}~\bibnamefont{Hu}}, \bibinfo{author}{\bibfnamefont{Z.}~\bibnamefont{Liu}}, \bibnamefont{et~al.}, \bibinfo{journal}{Nature} \textbf{\bibinfo{volume}{622}}, \bibinfo{pages}{74} (\bibinfo{year}{2023}), ISSN \bibinfo{issn}{1476-4687}, \bibinfo{note}{publisher: Nature Publishing Group}, \urlprefix\url{https://www.nature.com/articles/s41586-023-06536-0}.

\bibitem[{\citenamefont{Lu et~al.}(2024)\citenamefont{Lu, Han, Yao, Reddy, Yang, Seo, Watanabe, Taniguchi, Fu, and Ju}}]{lu:2024}
\bibinfo{author}{\bibfnamefont{Z.}~\bibnamefont{Lu}}, \bibinfo{author}{\bibfnamefont{T.}~\bibnamefont{Han}}, \bibinfo{author}{\bibfnamefont{Y.}~\bibnamefont{Yao}}, \bibinfo{author}{\bibfnamefont{A.~P.} \bibnamefont{Reddy}}, \bibinfo{author}{\bibfnamefont{J.}~\bibnamefont{Yang}}, \bibinfo{author}{\bibfnamefont{J.}~\bibnamefont{Seo}}, \bibinfo{author}{\bibfnamefont{K.}~\bibnamefont{Watanabe}}, \bibinfo{author}{\bibfnamefont{T.}~\bibnamefont{Taniguchi}}, \bibinfo{author}{\bibfnamefont{L.}~\bibnamefont{Fu}}, \bibnamefont{and} \bibinfo{author}{\bibfnamefont{L.}~\bibnamefont{Ju}}, \bibinfo{journal}{Nature} \textbf{\bibinfo{volume}{626}}, \bibinfo{pages}{759} (\bibinfo{year}{2024}), ISSN \bibinfo{issn}{1476-4687}, \bibinfo{note}{publisher: Nature Publishing Group}, \urlprefix\url{https://www.nature.com/articles/s41586-023-07010-7}.

\bibitem[{\citenamefont{Ren and Zhao}(2009)}]{ren:2009}
\bibinfo{author}{\bibfnamefont{Z.-A.} \bibnamefont{Ren}} \bibnamefont{and} \bibinfo{author}{\bibfnamefont{Z.-X.} \bibnamefont{Zhao}}, \bibinfo{journal}{Advanced Materials} \textbf{\bibinfo{volume}{21}}, \bibinfo{pages}{4584} (\bibinfo{year}{2009}), ISSN \bibinfo{issn}{1521-4095}, \urlprefix\url{https://onlinelibrary.wiley.com/doi/abs/10.1002/adma.200901049}.

\bibitem[{\citenamefont{Seibel et~al.}(2015)\citenamefont{Seibel, Schoop, Xie, Gibson, Webb, Fuccillo, Krizan, and Cava}}]{seibel:2015}
\bibinfo{author}{\bibfnamefont{E.~M.} \bibnamefont{Seibel}}, \bibinfo{author}{\bibfnamefont{L.~M.} \bibnamefont{Schoop}}, \bibinfo{author}{\bibfnamefont{W.}~\bibnamefont{Xie}}, \bibinfo{author}{\bibfnamefont{Q.~D.} \bibnamefont{Gibson}}, \bibinfo{author}{\bibfnamefont{J.~B.} \bibnamefont{Webb}}, \bibinfo{author}{\bibfnamefont{M.~K.} \bibnamefont{Fuccillo}}, \bibinfo{author}{\bibfnamefont{J.~W.} \bibnamefont{Krizan}}, \bibnamefont{and} \bibinfo{author}{\bibfnamefont{R.~J.} \bibnamefont{Cava}}, \bibinfo{journal}{Journal of the American Chemical Society} \textbf{\bibinfo{volume}{137}}, \bibinfo{pages}{1282} (\bibinfo{year}{2015}), ISSN \bibinfo{issn}{0002-7863}, \bibinfo{note}{publisher: American Chemical Society}, \urlprefix\url{https://doi.org/10.1021/ja511394q}.

\bibitem[{\citenamefont{Sinha et~al.}(2021)\citenamefont{Sinha, Vivanco, Wan, Siegler, Stewart, Pogue, Pressley, Berry, Wang, Johnson et~al.}}]{sinha:2021}
\bibinfo{author}{\bibfnamefont{M.}~\bibnamefont{Sinha}}, \bibinfo{author}{\bibfnamefont{H.~K.} \bibnamefont{Vivanco}}, \bibinfo{author}{\bibfnamefont{C.}~\bibnamefont{Wan}}, \bibinfo{author}{\bibfnamefont{M.~A.} \bibnamefont{Siegler}}, \bibinfo{author}{\bibfnamefont{V.~J.} \bibnamefont{Stewart}}, \bibinfo{author}{\bibfnamefont{E.~A.} \bibnamefont{Pogue}}, \bibinfo{author}{\bibfnamefont{L.~A.} \bibnamefont{Pressley}}, \bibinfo{author}{\bibfnamefont{T.}~\bibnamefont{Berry}}, \bibinfo{author}{\bibfnamefont{Z.}~\bibnamefont{Wang}}, \bibinfo{author}{\bibfnamefont{I.}~\bibnamefont{Johnson}}, \bibnamefont{et~al.}, \bibinfo{journal}{ACS Central Science} \textbf{\bibinfo{volume}{7}}, \bibinfo{pages}{1381} (\bibinfo{year}{2021}), ISSN \bibinfo{issn}{2374-7943}, \bibinfo{note}{publisher: American Chemical Society}, \urlprefix\url{https://doi.org/10.1021/acscentsci.1c00599}.

\bibitem[{\citenamefont{Hart et~al.}(2023)\citenamefont{Hart, Bhatt, Zhu, Han, Bianco, Li, Hynek, Schneeloch, Tao, Louca et~al.}}]{hart:2023}
\bibinfo{author}{\bibfnamefont{J.~L.} \bibnamefont{Hart}}, \bibinfo{author}{\bibfnamefont{L.}~\bibnamefont{Bhatt}}, \bibinfo{author}{\bibfnamefont{Y.}~\bibnamefont{Zhu}}, \bibinfo{author}{\bibfnamefont{M.-G.} \bibnamefont{Han}}, \bibinfo{author}{\bibfnamefont{E.}~\bibnamefont{Bianco}}, \bibinfo{author}{\bibfnamefont{S.}~\bibnamefont{Li}}, \bibinfo{author}{\bibfnamefont{D.~J.} \bibnamefont{Hynek}}, \bibinfo{author}{\bibfnamefont{J.~A.} \bibnamefont{Schneeloch}}, \bibinfo{author}{\bibfnamefont{Y.}~\bibnamefont{Tao}}, \bibinfo{author}{\bibfnamefont{D.}~\bibnamefont{Louca}}, \bibnamefont{et~al.}, \bibinfo{journal}{Nature Communications} \textbf{\bibinfo{volume}{14}}, \bibinfo{pages}{4803} (\bibinfo{year}{2023}), ISSN \bibinfo{issn}{2041-1723}, \bibinfo{note}{publisher: Nature Publishing Group}, \urlprefix\url{https://www.nature.com/articles/s41467-023-40528-y}.

\bibitem[{\citenamefont{Hu et~al.}(2024)\citenamefont{Hu, Qian, and Ni}}]{hu:2024}
\bibinfo{author}{\bibfnamefont{C.}~\bibnamefont{Hu}}, \bibinfo{author}{\bibfnamefont{T.}~\bibnamefont{Qian}}, \bibnamefont{and} \bibinfo{author}{\bibfnamefont{N.}~\bibnamefont{Ni}}, \bibinfo{journal}{National Science Review} \textbf{\bibinfo{volume}{11}}, \bibinfo{pages}{nwad282} (\bibinfo{year}{2024}), ISSN \bibinfo{issn}{2095-5138}, \urlprefix\url{https://doi.org/10.1093/nsr/nwad282}.

\bibitem[{\citenamefont{Wang and Steinfink}(1967)}]{wang:1967}
\bibinfo{author}{\bibfnamefont{R.}~\bibnamefont{Wang}} \bibnamefont{and} \bibinfo{author}{\bibfnamefont{H.}~\bibnamefont{Steinfink}}, \bibinfo{journal}{Inorganic Chemistry} \textbf{\bibinfo{volume}{6}}, \bibinfo{pages}{1685} (\bibinfo{year}{1967}), ISSN \bibinfo{issn}{0020-1669, 1520-510X}, \urlprefix\url{https://pubs.acs.org/doi/abs/10.1021/ic50055a017}.

\bibitem[{\citenamefont{Llanos et~al.}(2024)\citenamefont{Llanos, Campisi, Show, Kim, Dorrian, Salmani-Rezaie, Kioussis, and Falson}}]{llanos:2024}
\bibinfo{author}{\bibfnamefont{A.}~\bibnamefont{Llanos}}, \bibinfo{author}{\bibfnamefont{G.}~\bibnamefont{Campisi}}, \bibinfo{author}{\bibfnamefont{V.}~\bibnamefont{Show}}, \bibinfo{author}{\bibfnamefont{J.}~\bibnamefont{Kim}}, \bibinfo{author}{\bibfnamefont{R.}~\bibnamefont{Dorrian}}, \bibinfo{author}{\bibfnamefont{S.}~\bibnamefont{Salmani-Rezaie}}, \bibinfo{author}{\bibfnamefont{N.}~\bibnamefont{Kioussis}}, \bibnamefont{and} \bibinfo{author}{\bibfnamefont{J.}~\bibnamefont{Falson}}, \bibinfo{journal}{Nano Letters} \textbf{\bibinfo{volume}{24}}, \bibinfo{pages}{8518} (\bibinfo{year}{2024}), ISSN \bibinfo{issn}{1530-6984, 1530-6992}, \urlprefix\url{https://pubs.acs.org/doi/10.1021/acs.nanolett.4c01068}.

\bibitem[{\citenamefont{Hulliger and Schmelczer}(1978)}]{hulliger:1978}
\bibinfo{author}{\bibfnamefont{F.}~\bibnamefont{Hulliger}} \bibnamefont{and} \bibinfo{author}{\bibfnamefont{R.}~\bibnamefont{Schmelczer}}, \bibinfo{journal}{Journal of Solid State Chemistry} \textbf{\bibinfo{volume}{26}}, \bibinfo{pages}{389} (\bibinfo{year}{1978}), ISSN \bibinfo{issn}{00224596}, \urlprefix\url{https://linkinghub.elsevier.com/retrieve/pii/0022459678901743}.

\bibitem[{\citenamefont{Tremel and Hoffmann}(1987)}]{tremel:1987}
\bibinfo{author}{\bibfnamefont{W.}~\bibnamefont{Tremel}} \bibnamefont{and} \bibinfo{author}{\bibfnamefont{R.}~\bibnamefont{Hoffmann}}, \bibinfo{journal}{Journal of the American Chemical Society} \textbf{\bibinfo{volume}{109}}, \bibinfo{pages}{124} (\bibinfo{year}{1987}), ISSN \bibinfo{issn}{0002-7863, 1520-5126}, \urlprefix\url{https://pubs.acs.org/doi/abs/10.1021/ja00235a021}.

\bibitem[{\citenamefont{A.~Papoian and Hoffmann}(2000)}]{papoian:2000}
\bibinfo{author}{\bibfnamefont{G.}~\bibnamefont{A.~Papoian}} \bibnamefont{and} \bibinfo{author}{\bibfnamefont{R.}~\bibnamefont{Hoffmann}}, \bibinfo{journal}{Angewandte Chemie International Edition} \textbf{\bibinfo{volume}{39}}, \bibinfo{pages}{2408} (\bibinfo{year}{2000}), ISSN \bibinfo{issn}{1433-7851, 1521-3773}, \urlprefix\url{https://onlinelibrary.wiley.com/doi/10.1002/1521-3773(20000717)39:14<2408::AID-ANIE2408>3.0.CO;2-U}.

\bibitem[{\citenamefont{Ohno et~al.}(2021)\citenamefont{Ohno, Uchida, Kurihara, Minami, Nakazawa, Sato, Kriener, Hirayama, Miyake, Taguchi et~al.}}]{ohno:2021}
\bibinfo{author}{\bibfnamefont{M.}~\bibnamefont{Ohno}}, \bibinfo{author}{\bibfnamefont{M.}~\bibnamefont{Uchida}}, \bibinfo{author}{\bibfnamefont{R.}~\bibnamefont{Kurihara}}, \bibinfo{author}{\bibfnamefont{S.}~\bibnamefont{Minami}}, \bibinfo{author}{\bibfnamefont{Y.}~\bibnamefont{Nakazawa}}, \bibinfo{author}{\bibfnamefont{S.}~\bibnamefont{Sato}}, \bibinfo{author}{\bibfnamefont{M.}~\bibnamefont{Kriener}}, \bibinfo{author}{\bibfnamefont{M.}~\bibnamefont{Hirayama}}, \bibinfo{author}{\bibfnamefont{A.}~\bibnamefont{Miyake}}, \bibinfo{author}{\bibfnamefont{Y.}~\bibnamefont{Taguchi}}, \bibnamefont{et~al.}, \bibinfo{journal}{Physical Review B} \textbf{\bibinfo{volume}{103}}, \bibinfo{pages}{165144} (\bibinfo{year}{2021}), \bibinfo{note}{publisher: American Physical Society}, \urlprefix\url{https://link.aps.org/doi/10.1103/PhysRevB.103.165144}.

\bibitem[{\citenamefont{Kagayama et~al.}()\citenamefont{Kagayama, Oomi, and Bud}}]{kagayama:2000}
\bibinfo{author}{\bibfnamefont{T.}~\bibnamefont{Kagayama}}, \bibinfo{author}{\bibfnamefont{G.}~\bibnamefont{Oomi}}, \bibnamefont{and} \bibinfo{author}{\bibfnamefont{S.~L.} \bibnamefont{Bud}} (????).

\bibitem[{\citenamefont{Weinberger et~al.}(2023)\citenamefont{Weinberger, De~Podesta, Chen, Hodgson, and Grosche}}]{weinberger:2023}
\bibinfo{author}{\bibfnamefont{T.~I.} \bibnamefont{Weinberger}}, \bibinfo{author}{\bibfnamefont{C.~K.} \bibnamefont{De~Podesta}}, \bibinfo{author}{\bibfnamefont{J.}~\bibnamefont{Chen}}, \bibinfo{author}{\bibfnamefont{S.~A.} \bibnamefont{Hodgson}}, \bibnamefont{and} \bibinfo{author}{\bibfnamefont{F.~M.} \bibnamefont{Grosche}}, \bibinfo{journal}{SciPost Physics Proceedings} p. \bibinfo{pages}{018} (\bibinfo{year}{2023}), ISSN \bibinfo{issn}{2666-4003}, \urlprefix\url{https://scipost.org/10.21468/SciPostPhysProc.11.018}.

\bibitem[{\citenamefont{Squire et~al.}(2023)\citenamefont{Squire, Hodgson, Chen, Fedoseev, De~Podesta, Weinberger, Alireza, and Grosche}}]{squire:2023}
\bibinfo{author}{\bibfnamefont{O.~P.} \bibnamefont{Squire}}, \bibinfo{author}{\bibfnamefont{S.~A.} \bibnamefont{Hodgson}}, \bibinfo{author}{\bibfnamefont{J.}~\bibnamefont{Chen}}, \bibinfo{author}{\bibfnamefont{V.}~\bibnamefont{Fedoseev}}, \bibinfo{author}{\bibfnamefont{C.~K.} \bibnamefont{De~Podesta}}, \bibinfo{author}{\bibfnamefont{T.~I.} \bibnamefont{Weinberger}}, \bibinfo{author}{\bibfnamefont{P.~L.} \bibnamefont{Alireza}}, \bibnamefont{and} \bibinfo{author}{\bibfnamefont{F.~M.} \bibnamefont{Grosche}}, \bibinfo{journal}{Physical Review Letters} \textbf{\bibinfo{volume}{131}}, \bibinfo{pages}{026001} (\bibinfo{year}{2023}), ISSN \bibinfo{issn}{0031-9007, 1079-7114}, \urlprefix\url{https://link.aps.org/doi/10.1103/PhysRevLett.131.026001}.

\bibitem[{\citenamefont{Hitchcock et~al.}(2008)\citenamefont{Hitchcock, Lappert, Maron, and Protchenko}}]{hitchcock:2008}
\bibinfo{author}{\bibfnamefont{P.}~\bibnamefont{Hitchcock}}, \bibinfo{author}{\bibfnamefont{M.}~\bibnamefont{Lappert}}, \bibinfo{author}{\bibfnamefont{L.}~\bibnamefont{Maron}}, \bibnamefont{and} \bibinfo{author}{\bibfnamefont{A.}~\bibnamefont{Protchenko}}, \bibinfo{journal}{Angewandte Chemie International Edition} \textbf{\bibinfo{volume}{47}}, \bibinfo{pages}{1488} (\bibinfo{year}{2008}), ISSN \bibinfo{issn}{1433-7851, 1521-3773}, \urlprefix\url{https://onlinelibrary.wiley.com/doi/10.1002/anie.200704887}.

\bibitem[{\citenamefont{Pujar et~al.}(1995)\citenamefont{Pujar, Cawley, and Levine}}]{pujar:1995}
\bibinfo{author}{\bibfnamefont{V.~V.} \bibnamefont{Pujar}}, \bibinfo{author}{\bibfnamefont{J.~D.} \bibnamefont{Cawley}}, \bibnamefont{and} \bibinfo{author}{\bibfnamefont{S.~R.} \bibnamefont{Levine}}, \bibinfo{journal}{Journal of the American Ceramic Society} \textbf{\bibinfo{volume}{78}} (\bibinfo{year}{1995}), \bibinfo{note}{nTRS Author Affiliations: Case Western Reserve Univ., NTRS Document ID: 20010061714 NTRS Research Center: Glenn Research Center (GRC)}, \urlprefix\url{https://ntrs.nasa.gov/citations/20010061714}.

\bibitem[{\citenamefont{Kopp et~al.}(2012)\citenamefont{Kopp, Kaganer, Schwarzkopf, Waidick, Remmele, Kwasniewski, and Schmidbauer}}]{kopp:2012}
\bibinfo{author}{\bibfnamefont{V.~S.} \bibnamefont{Kopp}}, \bibinfo{author}{\bibfnamefont{V.~M.} \bibnamefont{Kaganer}}, \bibinfo{author}{\bibfnamefont{J.}~\bibnamefont{Schwarzkopf}}, \bibinfo{author}{\bibfnamefont{F.}~\bibnamefont{Waidick}}, \bibinfo{author}{\bibfnamefont{T.}~\bibnamefont{Remmele}}, \bibinfo{author}{\bibfnamefont{A.}~\bibnamefont{Kwasniewski}}, \bibnamefont{and} \bibinfo{author}{\bibfnamefont{M.}~\bibnamefont{Schmidbauer}}, \bibinfo{journal}{Acta Crystallographica Section A Foundations of Crystallography} \textbf{\bibinfo{volume}{68}}, \bibinfo{pages}{148} (\bibinfo{year}{2012}), ISSN \bibinfo{issn}{0108-7673, 1600-5724}, \urlprefix\url{https://journals.iucr.org/paper?S0108767311044874}.

\bibitem[{\citenamefont{Wagner et~al.}(2014)\citenamefont{Wagner, Baldini, Gogova, Schmidbauer, Schewski, Albrecht, Galazka, Klimm, and Fornari}}]{wagner:2014}
\bibinfo{author}{\bibfnamefont{G.}~\bibnamefont{Wagner}}, \bibinfo{author}{\bibfnamefont{M.}~\bibnamefont{Baldini}}, \bibinfo{author}{\bibfnamefont{D.}~\bibnamefont{Gogova}}, \bibinfo{author}{\bibfnamefont{M.}~\bibnamefont{Schmidbauer}}, \bibinfo{author}{\bibfnamefont{R.}~\bibnamefont{Schewski}}, \bibinfo{author}{\bibfnamefont{M.}~\bibnamefont{Albrecht}}, \bibinfo{author}{\bibfnamefont{Z.}~\bibnamefont{Galazka}}, \bibinfo{author}{\bibfnamefont{D.}~\bibnamefont{Klimm}}, \bibnamefont{and} \bibinfo{author}{\bibfnamefont{R.}~\bibnamefont{Fornari}}, \bibinfo{journal}{physica status solidi (a)} \textbf{\bibinfo{volume}{211}}, \bibinfo{pages}{27} (\bibinfo{year}{2014}), ISSN \bibinfo{issn}{1862-6319}, \bibinfo{note}{\_eprint: https://onlinelibrary.wiley.com/doi/pdf/10.1002/pssa.201330092}, \urlprefix\url{https://onlinelibrary.wiley.com/doi/ abs/10.1002/pssa.201330092}.

\bibitem[{\citenamefont{Canfield et~al.}(1991)\citenamefont{Canfield, Thompson, and Fisk}}]{canfield:1991}
\bibinfo{author}{\bibfnamefont{P.~C.} \bibnamefont{Canfield}}, \bibinfo{author}{\bibfnamefont{J.~D.} \bibnamefont{Thompson}}, \bibnamefont{and} \bibinfo{author}{\bibfnamefont{Z.}~\bibnamefont{Fisk}}, \bibinfo{journal}{Journal of Applied Physics} \textbf{\bibinfo{volume}{70}}, \bibinfo{pages}{5992} (\bibinfo{year}{1991}), ISSN \bibinfo{issn}{0021-8979}, \urlprefix\url{https://doi.org/10.1063/1.350071}.

\bibitem[{\citenamefont{Bud’ko et~al.}(1998)\citenamefont{Bud’ko, Canfield, Mielke, and Lacerda}}]{budko:1998}
\bibinfo{author}{\bibfnamefont{S.~L.} \bibnamefont{Bud’ko}}, \bibinfo{author}{\bibfnamefont{P.~C.} \bibnamefont{Canfield}}, \bibinfo{author}{\bibfnamefont{C.~H.} \bibnamefont{Mielke}}, \bibnamefont{and} \bibinfo{author}{\bibfnamefont{A.~H.} \bibnamefont{Lacerda}}, \bibinfo{journal}{Physical Review B} \textbf{\bibinfo{volume}{57}}, \bibinfo{pages}{13624} (\bibinfo{year}{1998}), ISSN \bibinfo{issn}{0163-1829, 1095-3795}, \urlprefix\url{https://link.aps.org/doi/10.1103/PhysRevB.57.13624}.

\bibitem[{\citenamefont{Coleman}(2015)}]{coleman:2015}
\bibinfo{author}{\bibfnamefont{P.}~\bibnamefont{Coleman}}, \emph{\bibinfo{title}{Heavy {Fermions} and the {Kondo} {Lattice}: a 21st {Century} {Perspective}}} (\bibinfo{year}{2015}), \bibinfo{note}{arXiv:1509.05769 [cond-mat]}, \urlprefix\url{http://arxiv.org/abs/1509.05769}.

\bibitem[{\citenamefont{Nishida et~al.}(2006)\citenamefont{Nishida, Tsuruta, and Miyake}}]{nishida:2006}
\bibinfo{author}{\bibfnamefont{Y.}~\bibnamefont{Nishida}}, \bibinfo{author}{\bibfnamefont{A.}~\bibnamefont{Tsuruta}}, \bibnamefont{and} \bibinfo{author}{\bibfnamefont{K.}~\bibnamefont{Miyake}}, \bibinfo{journal}{Journal of the Physical Society of Japan} \textbf{\bibinfo{volume}{75}}, \bibinfo{pages}{064706} (\bibinfo{year}{2006}), ISSN \bibinfo{issn}{0031-9015, 1347-4073}, \bibinfo{note}{arXiv:cond-mat/0606400}, \urlprefix\url{http://arxiv.org/abs/cond-mat/0606400}.

\bibitem[{\citenamefont{Zhang et~al.}(2024)\citenamefont{Zhang, Wu, Chen, Li, Smidman, Liu, Song, and Yuan}}]{zhang:2024}
\bibinfo{author}{\bibfnamefont{J.}~\bibnamefont{Zhang}}, \bibinfo{author}{\bibfnamefont{J.}~\bibnamefont{Wu}}, \bibinfo{author}{\bibfnamefont{Y.}~\bibnamefont{Chen}}, \bibinfo{author}{\bibfnamefont{R.}~\bibnamefont{Li}}, \bibinfo{author}{\bibfnamefont{M.}~\bibnamefont{Smidman}}, \bibinfo{author}{\bibfnamefont{Y.}~\bibnamefont{Liu}}, \bibinfo{author}{\bibfnamefont{Y.}~\bibnamefont{Song}}, \bibnamefont{and} \bibinfo{author}{\bibfnamefont{H.}~\bibnamefont{Yuan}}, \bibinfo{journal}{Chinese Physics Letters} \textbf{\bibinfo{volume}{41}}, \bibinfo{pages}{127304} (\bibinfo{year}{2024}), ISSN \bibinfo{issn}{0256-307X, 1741-3540}, \bibinfo{note}{arXiv:2501.07889 [cond-mat]}, \urlprefix\url{http://arxiv.org/abs/2501.07889}.

\bibitem[{\citenamefont{Posey et~al.}(2024)\citenamefont{Posey, Turkel, Rezaee, Devarakonda, Kundu, Ong, Thinel, Chica, Vitalone, Jing et~al.}}]{posey:2024}
\bibinfo{author}{\bibfnamefont{V.~A.} \bibnamefont{Posey}}, \bibinfo{author}{\bibfnamefont{S.}~\bibnamefont{Turkel}}, \bibinfo{author}{\bibfnamefont{M.}~\bibnamefont{Rezaee}}, \bibinfo{author}{\bibfnamefont{A.}~\bibnamefont{Devarakonda}}, \bibinfo{author}{\bibfnamefont{A.~K.} \bibnamefont{Kundu}}, \bibinfo{author}{\bibfnamefont{C.~S.} \bibnamefont{Ong}}, \bibinfo{author}{\bibfnamefont{M.}~\bibnamefont{Thinel}}, \bibinfo{author}{\bibfnamefont{D.~G.} \bibnamefont{Chica}}, \bibinfo{author}{\bibfnamefont{R.~A.} \bibnamefont{Vitalone}}, \bibinfo{author}{\bibfnamefont{R.}~\bibnamefont{Jing}}, \bibnamefont{et~al.}, \bibinfo{journal}{Nature} \textbf{\bibinfo{volume}{625}}, \bibinfo{pages}{483} (\bibinfo{year}{2024}), ISSN \bibinfo{issn}{0028-0836, 1476-4687}, \urlprefix\url{https://www.nature.com/articles/s41586-023-06868-x}.

\bibitem[{\citenamefont{Zhang et~al.}(2017)\citenamefont{Zhang, Zhu, Hu, Tan, Xie, Feng, Qin, Zhang, Liu, Song et~al.}}]{zhang:2017}
\bibinfo{author}{\bibfnamefont{Y.}~\bibnamefont{Zhang}}, \bibinfo{author}{\bibfnamefont{X.}~\bibnamefont{Zhu}}, \bibinfo{author}{\bibfnamefont{B.}~\bibnamefont{Hu}}, \bibinfo{author}{\bibfnamefont{S.}~\bibnamefont{Tan}}, \bibinfo{author}{\bibfnamefont{D.}~\bibnamefont{Xie}}, \bibinfo{author}{\bibfnamefont{W.}~\bibnamefont{Feng}}, \bibinfo{author}{\bibfnamefont{L.}~\bibnamefont{Qin}}, \bibinfo{author}{\bibfnamefont{W.}~\bibnamefont{Zhang}}, \bibinfo{author}{\bibfnamefont{Y.}~\bibnamefont{Liu}}, \bibinfo{author}{\bibfnamefont{H.}~\bibnamefont{Song}}, \bibnamefont{et~al.}, \bibinfo{journal}{Chinese Physics B} \textbf{\bibinfo{volume}{26}}, \bibinfo{pages}{067102} (\bibinfo{year}{2017}), ISSN \bibinfo{issn}{1674-1056}, \urlprefix\url{https://iopscience.iop.org/article/10.1088/1674-1056/26/6/067102}.

\bibitem[{\citenamefont{Trainer et~al.}(2021)\citenamefont{Trainer, Abel, Bud'ko, Canfield, and Wahl}}]{trainer:2021}
\bibinfo{author}{\bibfnamefont{C.}~\bibnamefont{Trainer}}, \bibinfo{author}{\bibfnamefont{C.}~\bibnamefont{Abel}}, \bibinfo{author}{\bibfnamefont{S.~L.} \bibnamefont{Bud'ko}}, \bibinfo{author}{\bibfnamefont{P.~C.} \bibnamefont{Canfield}}, \bibnamefont{and} \bibinfo{author}{\bibfnamefont{P.}~\bibnamefont{Wahl}}, \bibinfo{journal}{Physical Review B} \textbf{\bibinfo{volume}{104}}, \bibinfo{pages}{205134} (\bibinfo{year}{2021}), ISSN \bibinfo{issn}{2469-9950, 2469-9969}, \urlprefix\url{https://link.aps.org/doi/10.1103/PhysRevB.104.205134}.

\bibitem[{\citenamefont{Nair et~al.}(2012)\citenamefont{Nair, Wirth, Friedemann, Steglich, Si, and Schofield}}]{nair:2012}
\bibinfo{author}{\bibfnamefont{S.}~\bibnamefont{Nair}}, \bibinfo{author}{\bibfnamefont{S.}~\bibnamefont{Wirth}}, \bibinfo{author}{\bibfnamefont{S.}~\bibnamefont{Friedemann}}, \bibinfo{author}{\bibfnamefont{F.}~\bibnamefont{Steglich}}, \bibinfo{author}{\bibfnamefont{Q.}~\bibnamefont{Si}}, \bibnamefont{and} \bibinfo{author}{\bibfnamefont{A.~J.} \bibnamefont{Schofield}}, \bibinfo{journal}{Advances in Physics} \textbf{\bibinfo{volume}{61}}, \bibinfo{pages}{583} (\bibinfo{year}{2012}), ISSN \bibinfo{issn}{0001-8732, 1460-6976}, \bibinfo{note}{arXiv:1210.0809 [cond-mat]}, \urlprefix\url{http://arxiv.org/abs/1210.0809}.

\bibitem[{\citenamefont{Holter and Adrian}(1985)}]{holter:1985}
\bibinfo{author}{\bibfnamefont{G.}~\bibnamefont{Holter}} \bibnamefont{and} \bibinfo{author}{\bibfnamefont{H.}~\bibnamefont{Adrian}}, \bibinfo{journal}{Solid State Communications} \textbf{\bibinfo{volume}{58}} (\bibinfo{year}{1985}).

\bibitem[{\citenamefont{Fert and Levy}(1987)}]{fert:1987}
\bibinfo{author}{\bibfnamefont{A.}~\bibnamefont{Fert}} \bibnamefont{and} \bibinfo{author}{\bibfnamefont{P.~M.} \bibnamefont{Levy}}, \bibinfo{journal}{Physical Review B} \textbf{\bibinfo{volume}{36}}, \bibinfo{pages}{1907} (\bibinfo{year}{1987}), ISSN \bibinfo{issn}{0163-1829}, \urlprefix\url{https://link.aps.org/doi/10.1103/PhysRevB.36.1907}.

\bibitem[{\citenamefont{Hamzic et~al.}(1988)\citenamefont{Hamzic, Fert, Miljak, and Horn}}]{hamzic:1988}
\bibinfo{author}{\bibfnamefont{A.}~\bibnamefont{Hamzic}}, \bibinfo{author}{\bibfnamefont{A.}~\bibnamefont{Fert}}, \bibinfo{author}{\bibfnamefont{M.}~\bibnamefont{Miljak}}, \bibnamefont{and} \bibinfo{author}{\bibfnamefont{S.}~\bibnamefont{Horn}}, \bibinfo{journal}{Physical Review B} \textbf{\bibinfo{volume}{38}}, \bibinfo{pages}{7141} (\bibinfo{year}{1988}), ISSN \bibinfo{issn}{0163-1829}, \urlprefix\url{https://link.aps.org/doi/10.1103/PhysRevB.38.7141}.

\bibitem[{\citenamefont{Kresse and Furthm\"uller}(1996{\natexlab{a}})}]{Kresse1996}
\bibinfo{author}{\bibfnamefont{G.}~\bibnamefont{Kresse}} \bibnamefont{and} \bibinfo{author}{\bibfnamefont{J.}~\bibnamefont{Furthm\"uller}}, \bibinfo{journal}{Phys. Rev. B} \textbf{\bibinfo{volume}{54}}, \bibinfo{pages}{11169} (\bibinfo{year}{1996}{\natexlab{a}}), \urlprefix\url{https://link.aps.org/doi/10.1103/PhysRevB.54.11169}.

\bibitem[{\citenamefont{Kresse and Furthm\"uller}(1996{\natexlab{b}})}]{Kresse1996b}
\bibinfo{author}{\bibfnamefont{G.}~\bibnamefont{Kresse}} \bibnamefont{and} \bibinfo{author}{\bibfnamefont{J.}~\bibnamefont{Furthm\"uller}}, \bibinfo{journal}{Computational Materials Science} \textbf{\bibinfo{volume}{6}}, \bibinfo{pages}{15 } (\bibinfo{year}{1996}{\natexlab{b}}), ISSN \bibinfo{issn}{0927-0256}, \urlprefix\url{http://www.sciencedirect.com/science/article/pii/0927025696000080}.

\bibitem[{\citenamefont{Bl\"ochl}(1994)}]{Blochl1994}
\bibinfo{author}{\bibfnamefont{P.~E.} \bibnamefont{Bl\"ochl}}, \bibinfo{journal}{Phys. Rev. B} \textbf{\bibinfo{volume}{50}}, \bibinfo{pages}{17953} (\bibinfo{year}{1994}), \urlprefix\url{https://link.aps.org/doi/10.1103/PhysRevB.50.17953}.

\bibitem[{\citenamefont{Kresse and Joubert}(1999)}]{Kresse1999}
\bibinfo{author}{\bibfnamefont{G.}~\bibnamefont{Kresse}} \bibnamefont{and} \bibinfo{author}{\bibfnamefont{D.}~\bibnamefont{Joubert}}, \bibinfo{journal}{Phys. Rev. B} \textbf{\bibinfo{volume}{59}}, \bibinfo{pages}{1758} (\bibinfo{year}{1999}), \urlprefix\url{https://link.aps.org/doi/10.1103/PhysRevB.59.1758}.

\bibitem[{\citenamefont{Perdew et~al.}(1996)\citenamefont{Perdew, Burke, and Ernzerhof}}]{PBE}
\bibinfo{author}{\bibfnamefont{J.~P.} \bibnamefont{Perdew}}, \bibinfo{author}{\bibfnamefont{K.}~\bibnamefont{Burke}}, \bibnamefont{and} \bibinfo{author}{\bibfnamefont{M.}~\bibnamefont{Ernzerhof}}, \bibinfo{journal}{Phys. Rev. Lett.} \textbf{\bibinfo{volume}{77}}, \bibinfo{pages}{3865} (\bibinfo{year}{1996}), \urlprefix\url{https://link.aps.org/doi/10.1103/PhysRevLett.77.3865}.

\bibitem[{\citenamefont{Togo et~al.}(2023)\citenamefont{Togo, Chaput, Tadano, and Tanaka}}]{Togo_2023}
\bibinfo{author}{\bibfnamefont{A.}~\bibnamefont{Togo}}, \bibinfo{author}{\bibfnamefont{L.}~\bibnamefont{Chaput}}, \bibinfo{author}{\bibfnamefont{T.}~\bibnamefont{Tadano}}, \bibnamefont{and} \bibinfo{author}{\bibfnamefont{I.}~\bibnamefont{Tanaka}}, \bibinfo{journal}{Journal of Physics: Condensed Matter} \textbf{\bibinfo{volume}{35}}, \bibinfo{pages}{353001} (\bibinfo{year}{2023}), \urlprefix\url{https://dx.doi.org/10.1088/1361-648X/acd831}.

\bibitem[{\citenamefont{Togo}(2023)}]{Togo_2023b}
\bibinfo{author}{\bibfnamefont{A.}~\bibnamefont{Togo}}, \bibinfo{journal}{Journal of the Physical Society of Japan} \textbf{\bibinfo{volume}{92}}, \bibinfo{pages}{012001} (\bibinfo{year}{2023}), \eprint{https://doi.org/10.7566/JPSJ.92.012001}, \urlprefix\url{https://doi.org/10.7566/JPSJ.92.012001}.

\bibitem[{\citenamefont{Flack}(2016)}]{tables}
\bibinfo{author}{\bibfnamefont{D.~H.} \bibnamefont{Flack}} (\bibinfo{publisher}{International Union for Crystallography}, \bibinfo{year}{2016}), vol.~\bibinfo{volume}{A}, chap. \bibinfo{chapter}{1.6}, pp. \bibinfo{pages}{114--128}.

\end{thebibliography}

\clearpage

\renewcommand{\thefigure}{S\arabic{figure}}

\setcounter{figure}{0}

\subsection*{Supplementary Data}

\section{Experimental Section}

\textit{Density Functional Theory}:
First principles calculations were performed with the Vienna $ab$ $initio$ simulation package ({\small VASP}).\cite{Kresse1996,Kresse1996b} The pseudopotential and wave functions are represented within the projector augmented wave method.\cite{Blochl1994,Kresse1999} The Ce-$f$ band is treated as a core state and the valence configuration, $5s^25p^66s^25d^1$ is identical with La. It is $5s^25p^3$ for Sb. The exchange correlation functional is described by GGA-PBE.\cite{PBE} The plane wave cutoff energy is set 500 eV and a $16 \times 16 \times 6$ $k$-point mesh is used in the momentum space sampling. Crystal structures are optimized within energy convergence criteria of $10^{-5}$ eV. The dynamical matrices are calculated by using {\small VASP}-{\small PHONOPY}\cite{Togo_2023,Togo_2023b} interface using finite displacement method in
$2 \times 2 \times 1$ supercell (16 $R$ and 32 Sb atoms) structures with $8 \times 8 \times 6$ $k$-point mesh. The temperature dependent free energy accounting for phonon entropy is written as,
\begin{equation}
	F(T) = E_0 + \frac{1}{2}\sum_{\bm{q}n}{\hbar \omega_{\bm{q}n}} + \sum_{\bm{q}n}{k_B T \ln\left(1-e^{-\hbar \omega_{\bm{q}n}/k_B T}\right)},
\end{equation}
where $E_0$ is the total energy with frozen ions, $\omega_{{\bm q}n}$ is the phonon frequency of $n$-th mode with a wave vector $\bm{q}$. The second and third terms are the phonon zero energy and entropy contributions, respectively.

\textit{Thin Film Growth}:
\ce{CeSb2} films were prepared using MBE according to the recipe reported by \ce{LaSb2} in Ref.[\onlinecite{llanos:2024}]. Here we modify the growth recipe by excluding the low-temperature buffer layer and growing 30-40 nm films at a fixed growth temperature directly on annealed MgO(001) substrates. We also vary the Sb:Ce flux ratio and the flux of co-evaporated La to explore the phase stability of the Ce-La-Sb system across multiple tunable parameters.

\textit{X-Ray Diffraction}:
Crystallographic data was obtained with a Rigaku Smartlab X-ray diffractometer using a 2-bounce Ge (220) monochromator and Cu-K$_{\alpha1}$ radiation. Identification of phase, film thickness, and epitaxial orientation was performed through out-of-plane $\theta/2\theta$ scans. Asymmetric reciprocal-space mapping was used to further identify the present crystal structures. Rocking curves and $\varphi$-scans are reported in the Supplementary Materials as measures of the films' mosaicity and in-plane crystalline order. 

\textit{Electronic Transport}:
Electrical data were collected using a Quantum Design Dynacool PPMS with a base temperature of $\approx 1.7$ K and maximum magnetic field of either $9$ or $14$ T. Samples were cut into approximately square-shaped pieces ($\approx 1$ - $2$mm side lengths) and wire-bonded into a four-point van der Pauw geometry. Numerically solving the van der Pauw equation using the measured $R_{xx}$ and $R_{yy}$ yields a sheet resistance, which we report. For measurements of magnetoresistance and Hall effect, the default PPMS rotator assembly was used to orient the applied field either parallel to the film surface ($H\parallel$ab) or perpendicular to the film ($H\parallel$c), as identified by an initial high-field rotation sweep. Magnetoresistance traces were obtained by symmetrizing the sheet resistance data with respect to the applied field, carefully symmetrizing each data point with its time-reversed counterpart (up-sweep symmetrized against down-sweep) to preserve hysteretic effects. Similarly, \textit{anti}-symmetrization of the diagonal voltage data under $H\parallel$c was employed to isolate the Hall resistivity.

\newpage
\begin{table*}[h]
	\begin{ruledtabular}
		\begin{tabular}{r|c|ccccccc}
			System &  Space Group\; & $a$ (\AA) & $b$ (\AA) & $c$ (\AA) & $\alpha$ ($^\circ$) & $\beta$ ($^\circ$) & $\gamma$ ($^\circ$) \\
			\hline
			\ce{LaSb2}:\; Sm-type & $Cmca$ (64) \; &  4.471 & 4.471 & 18.63 & 90 & 90 & 91.1 \\
            Yb-mono& $A2/m$ (12) \; & 4.562 & 4.483 & 17.71 & 90 & 86.3 & 90 \\

            \hline
            \ce{CeSb2}:\; Sm-type & $Cmca$ (64) \; &  4.452 & 4.452 & 18.67 & 90 & 90 & 91.2 \\
            Yb-mono& $A2/m$ (12) \; & 4.559 & 4.463 & 17.75 & 90 & 86.0 & 90 \\
            \hline\hline
            \ce{CeSb2}, experiment: & & & & & & & \\
            Sm-type\cite{wang:1967} & $Cmca$ (64) \; &  6.295 & 6.124 & 18.21 & 90 & 90 & 90 \\
            Yb-mono (predicted)\cite{llanos:2024} & $A2/m$ (12) \; & 4.458 & 4.381 & 17.31 & 90 & 86.3 & 90 \\

            \hline
            This work, low-T & &  6.27 & 6.16 & 18.16 & 90 & 90 & 90 \\
            This work, high-T & &  4.49 & 4.39 & 17.38 & 90 & 86.1 & 90 \\
		\end{tabular}
	\end{ruledtabular}
    
    \caption{\label{tab:bulk}
    Structural parameters of both stacking configurations of \ce{CeSb2}. Top: parameters obtained from and used in first-principles DFT calculations. Note that the Sm-type structure is here defined differently than in the text as a base-centered orthorhombic cell, where the lattice constants a and b are equivalent to the distance from the corner to the center of the conventional cell base. Bottom: expected parameters as found from bulk studies (Sm-type) and scaling those of Yb-mono \ce{LaSb2} proportional to the lanthanide contraction observed in the bulk Sm-type ($\approx 2.3\%$), compared to parameters measured in this work as reported in the main text. A few-percent discrepancy between DFT-obtained and experimental lattice parameters is expected.
    }
\end{table*}
\begin{figure}
    \centering
    \includegraphics[width=160mm]{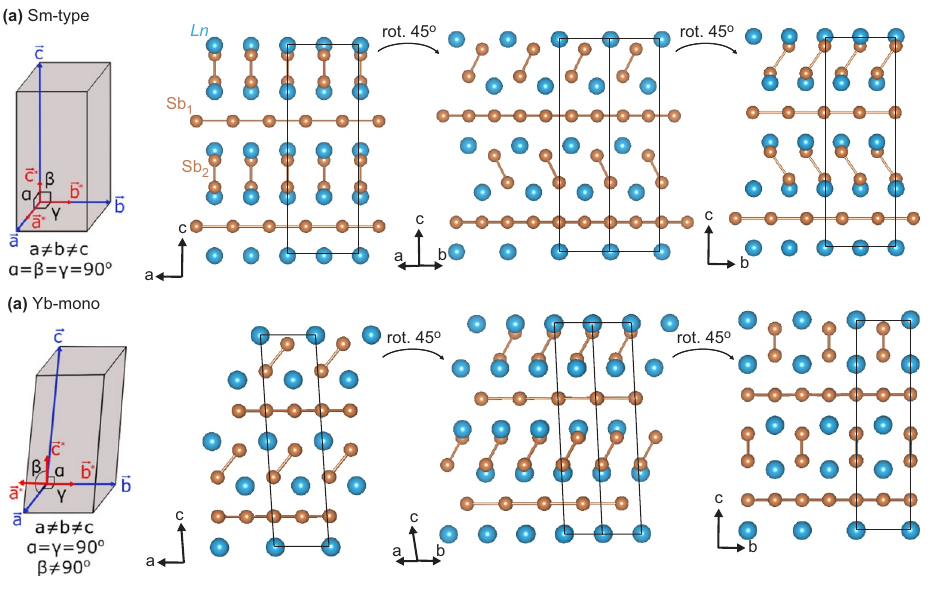}
    \caption{Crystal structure of the (a) Sm-type and (b) Yb-mono stacking configurations along various in-plane viewing angles.}
    \label{FigS1}
\end{figure}

\newpage

\begin{figure}[h]
    \centering
    \includegraphics[width=160mm]{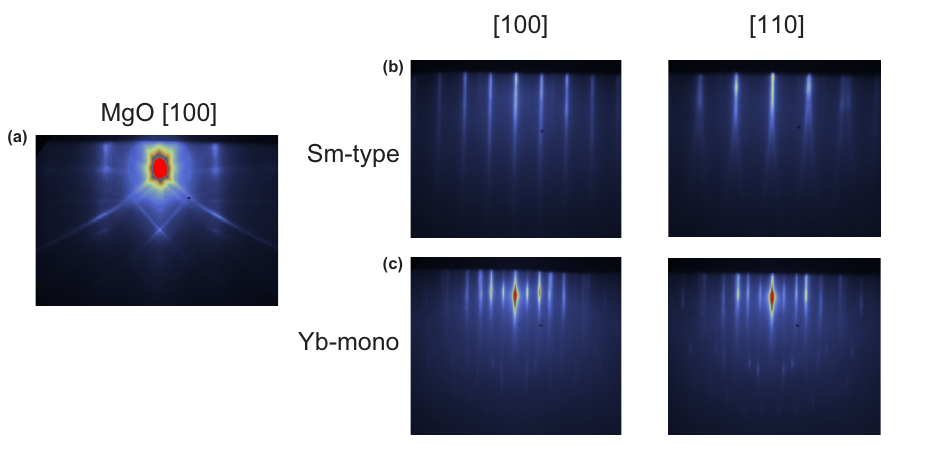}
    \caption{In-situ RHEED diffraction patterns of (a) an annealed MgO substrate and \ce{CeSb2} films grown in the (a) Sm-type and (c) Yb-mono polymorphs. The higher growth temperature of the Yb-monoclinic films results in better surface crystallinity (seen by the sharpness and brightness of the RHEED streaks) however a more prominent 45$^\circ$-rotated domain as seen when comparing the [100] and [110]-incident RHEED images.}
    \label{FigS2}
\end{figure}

\newpage

\begin{figure}[h]
    \centering
    \includegraphics[width=160mm]{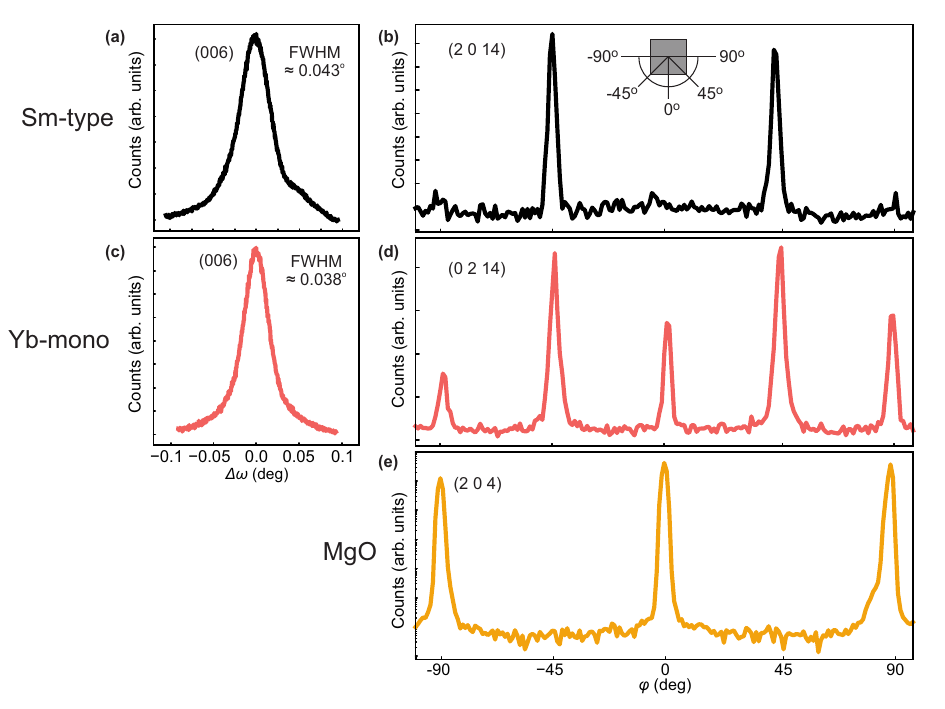}
    \caption{Rocking curves and $\varphi$-scans for (a-b) Sm-type and (c-d) Yb-mono films. Rocking curves taken for the (006) peaks of both polymorphs are very sharp with full widths at half maximum FWHM of $<$ 0.05$^\circ$, ensuring high crystallinity of the films. $\varphi$-scans of the {2 0 14} peaks reveals a four-fold symmetry and dominant [100]$_{\text{CeSb}_2}\parallel$ [110]$_{\text{MgO}}$ epitaxial register for both films, however the higher-$T_\mathrm{g}$ Yb-mono film also shows 45$^\circ$-rotated domains consistent with our observations in RHEED. (e) $\varphi$-scan of the MgO (204) peak for reference.}
    \label{FigS3}
\end{figure}

\newpage
\begin{figure}[h]
    \centering
    \includegraphics[width=160mm]{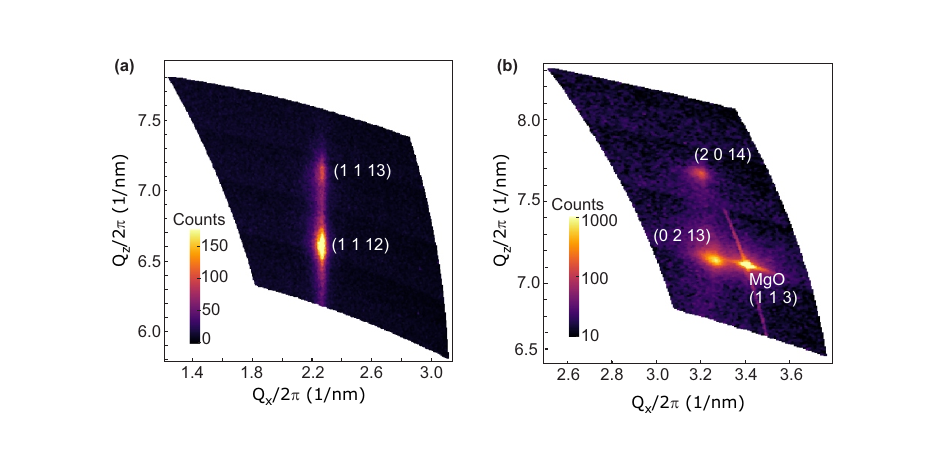}
    \caption{Additional RSM scans of the Sm-type structure. Consistent with extinction rules (a) $h+k = 2n$ for $h,k,l \neq 0$ and (b) $h,\ell=2n,\,k=0$ and $k=2n,\,h=0$. All rules agree with those of the $Cmca$ space group.\cite{tables} }
    \label{FigS4}
\end{figure}

\newpage

\begin{figure}[h]
    \centering
    \includegraphics[width=120mm]{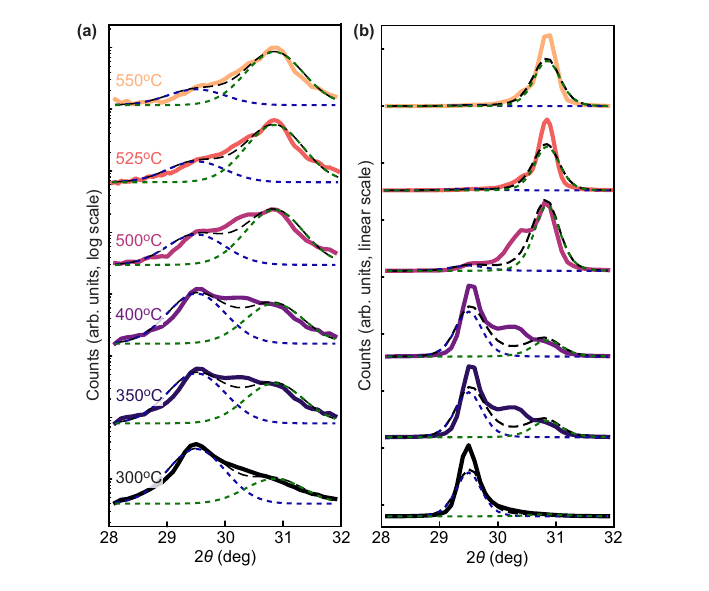}
    \caption{Two-gaussian fits to the (006) Bragg peak across multiple growth temperatures. The peak positions were fixed according to observed $c$-axis lattice constants of the two polymorphs and the gaussians' widths and amplitudes were free to vary. The relative amplitude of the two gaussians was recorded as a percent-difference and reported throughout the work as a measure of phase purity. (a) Plotted on a log scale as the fits were performed. (b) On a linear scale which provides a better visualization of the phase purity of the $300^o$C and $525$-$550^o$C films.}
    \label{FigS5}
\end{figure}

\newpage

\begin{figure}[h]
    \centering
    \includegraphics[width=160mm]{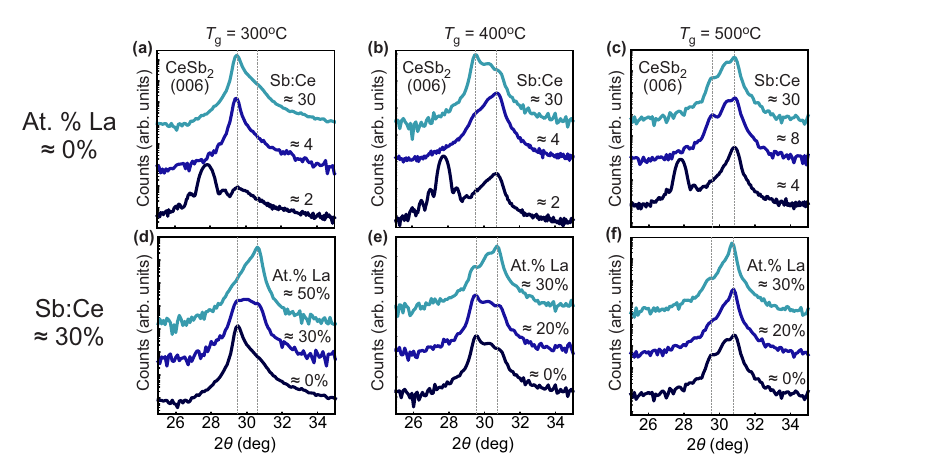}
    \caption{(006) Bragg peaks for selected films on the phase diagrams reported in the main text's Figure 3a (a-c) and 3b (d-f). When lowering the Sb:Ce flux ratio, all growth temperatures eventually display prominent CeSb impurity peaks at sufficient levels of Sb deficiency. Meanwhile, a clear hand-off from Sm-type to Yb-mono is seen as the higher-$T_g$ films approach Sb-deficiency from above. A similar trade-off can be seen in (d-f) as the \%-substitution of La for Ce is increased. Two-gaussian fits to these lineshapes (excluding the CeSb impurity peak) provided the relative peak intensity data reported in Figure 3 in the main text.}
    \label{FigS6}
\end{figure}

\newpage

\begin{figure}[h]
    \centering
    \includegraphics[width=160mm]{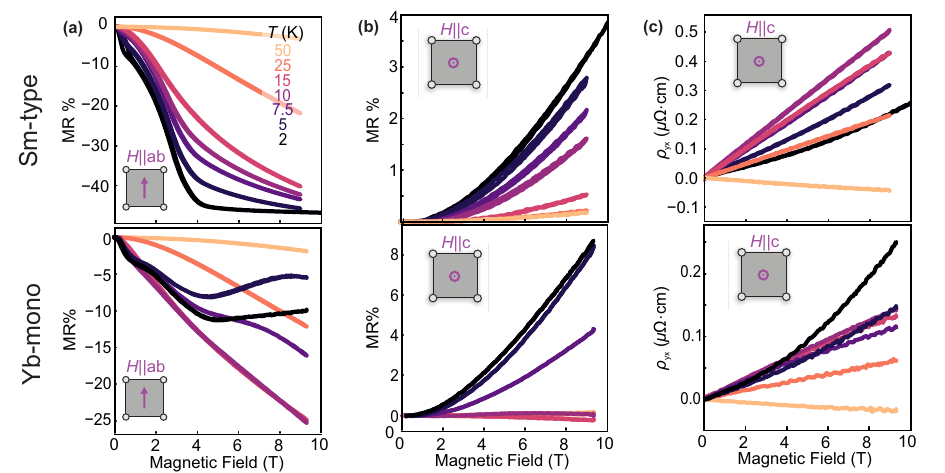}
    \caption{Temperature series of magneto-transport behavior for the Yb-monoclinic and Sm-type polymorphs. (a) $H\parallel ab$ longitudinal magnetoresistance, (b) $H\parallel c$ out-of-plane longitudinal magnetoresistance, and (c) Hall resistivity.}
    \label{FigS7}
\end{figure}

\end{document}